\RequirePackage{ifpdf}
\ifpdf 
\documentclass[pdftex]{sigma}
\else
\documentclass{sigma}
\fi

\begin{document}

\allowdisplaybreaks

\renewcommand{\PaperNumber}{056}

\FirstPageHeading

\renewcommand{\thefootnote}{$\star$}

\ShortArticleName{Tridiagonal Symmetries of Models of Nonequilibrium
Physics}

\ArticleName{Tridiagonal Symmetries of Models\\ of Nonequilibrium
Physics\footnote{This
paper is a contribution to the Proceedings of the Seventh
International Conference ``Symmetry in Nonlinear Mathematical
Physics'' (June 24--30, 2007, Kyiv, Ukraine). The full collection
is available at
\href{http://www.emis.de/journals/SIGMA/symmetry2007.html}{http://www.emis.de/journals/SIGMA/symmetry2007.html}}}

\Author{Boyka ANEVA}

\AuthorNameForHeading{B. Aneva}

\Address{Institute for Nuclear Research and Nuclear Energy,
Bulgarian Academy of Sciences,\\ 72 Tzarigradsko chaussee, 1784
Sof\/ia, Bulgaria}
\Email{\href{mailto:blan@inrne.bas.bg}{blan@inrne.bas.bg}}

\ArticleDates{Received March 03, 2008, in f\/inal form July 14,
2008; Published online July 28, 2008}

\Abstract{We study the boundary symmetries of models of
nonequilibrium physics where the steady state behaviour strongly
depends on the boundary rates.  Within the matrix product state
approach to many-body systems  the physics is described  in terms
of matrices def\/ining a noncommutative space with a quantum group
symmetry. Boundary processes lead to a reduction of the bulk
symmetry. We argue that the boundary operators of an interacting
system with simple exclusion generate a tridiagonal algebra whose
irreducible representations are expressed in terms of the
Askey--Wilson polynomials. We show that the boundary algebras of
the symmetric and the totally asymmetric processes are the proper
limits of the partially asymmetric ones. In all three type of
processes the tridiagonal algebra arises as a symmetry of the
boundary problem and allows for the exact solvability of  the
model.}


\Classification{82C10; 60J60; 17B80}

\section{Introduction}

Out of the rich variety of phenomena in nature the most
interesting occur in nonequilibrium conditions and their complex
behaviour is far from being well understood. The study of
nonequilibrium phenomena is of current interest and  the way to
describe the general characteristics of a system out of
equilibrium goes through the analysis of mathematical models. Such
models have to be simple enough, still they  must be physically
signif\/icant to exhibit the structure of the complex phenomena.
Stochastic interacting particle systems \cite{nneq, schu1, kul2}
received a lot of attention. Among these, the asymmetric simple
exclusion process (ASEP) has become a paradigm in nonequilibrium
physics due to its simplicity, rich behaviour and wide range of
applicability.
  Introduced originally as a simplif\/ied model of one
dimensional transport for phenomena like kinetics of
biopolymerization \cite{mac}, it has found applications from
traf\/f\/ic f\/low \cite{chow, helb}, to interface growth \cite{spo},
shock formation, hydrodynamic systems obeying the noisy Burger
equation, problems of sequence alignment in biology \cite{bu}.

At large time the ASEP exhibits relaxation to a steady state, and
even after the relaxation it has a nonvanishing current. An
intriguing feature is the occurrence of boundary induced phase
transitions \cite{schu1} and the fact that the bulk properties
depend strongly on the boundary rates.

The asymmetric exclusion process is an exactly solvable model of a
lattice dif\/fusion system of particles interacting with a hard core
exclusion, i.e.\ the lattice site can be either empty or occupied
by a particle. As a stochastic process it is described in terms of
a probability distribution $P(s_m, t)$ of a stochastic variable
$s_m = 0, 1$ at a site $m = 1,2,\dots,L$ of a linear chain. A state
on the lattice at a time $t$ is determined by the set of
occupation numbers $s_1, s_2,\dots,s_L$ and a transition to another
conf\/iguration $s'$ during an inf\/initesimal time step $dt$ is given
by the probability $\Gamma(s, s')dt$.  With the restriction of
dynamics that changes of conf\/iguration can only occur at two
adjacent sites, the rates for such changes depend on these sites
only. The two-site rates $\Gamma (s,s') \equiv \Gamma ^{s_m,
s_{m+1}}_{s'_m,s'_{m+1}} = \Gamma ^{ik}_{jl}$, $i,j,k,l =0, 1$ are
assumed to be independent from the position in the bulk. Due to
probability conservation $\Gamma (s,s) = -\sum_{s'\neq s} \Gamma
(s',s)$. For dif\/fusion processes the transition rate matrix
becomes simply $\Gamma ^{ik}_{ki}=g_{ik}$. At the boundaries, i.e.\
sites~$1$ and~$L$ additional processes can take place with rates
$L_i^j$ and $R_i^j$ ($i,j =0, 1$). In the set of occupation
numbers $(s_1,s_2,\dots,s_L)$ specifying a conf\/iguration of the
system $s_i=0$ if a site $i$ is empty, $s_i=1$ if there is a
particle at a site $i$. Particles hop to the left with probability
$g_{01}dt$ and to the right with probability $g_{10}dt$. The event
of exchange happens if out of two adjacent sites one is a vacancy
and the other is occupied by a particle. The symmetric simple
exclusion process is known as the lattice gas model of particle
hopping between nearest-neighbour sites with a constant rate
$g_{01}=g_{10}=g$. The partially asymmetric simple exclusion
process with hopping in a preferred direction is the driven
dif\/fusive lattice gas of particles moving under the action of an
external f\/ield. The process is totally asymmetric if all jumps
occur in one direction only, and partially asymmetric if there is
a dif\/ferent non-zero probability of both left and right hopping.
The number of particles in the bulk is conserved and this is the
case of periodic boundary conditions. In the case of open systems,
the lattice gas is coupled to external reservoirs of particles of
f\/ixed density. The most interesting examples  (see~\cite{evbl} and references therein
for a review) are phase transitions inducing boundary processes
\cite{schu1, kru} when a particle is added with probability
$\alpha dt$ and/or removed with probability $\gamma dt$ at the
left end of the chain, and it is removed with
 probability
$\beta dt$ and/or added with probability $\delta dt$
 at the right
 end of the chain.

The time evolution of the model is governed by the master equation
for the probability distribution of the stochastic system
 $\frac {dP(s,t)}{dt} = \sum _{s'} \Gamma (s,s')P(s',t)$.
It can be mapped to a Schr\"{o}dinger equation in imaginary time
for a quantum Hamiltonian with nearest-neighbour interaction in
the bulk and single-site boundary terms $\frac {dP(t)}{dt}=-HP(t)$,
where $ H= \sum_j H_{j,j+1}+ H^{(L)} + H^{(R)}$ and the ground
state of the Hamiltonian, in general non-Hermitian, corresponds to
the steady state of the stochastic dynamics where all
probabilities are stationary. The mapping $\Gamma =- \sqrt q
U^{-1}_{\mu}H_{XXZ}U_{\mu}$, where
\[
U_{\mu}=\bigotimes_{i=1}^{L}\begin{pmatrix} 1 & 0     \\
                                              0 & \mu(\sqrt q)^{i-1}
\end{pmatrix}
\]
(see \cite{ess} and \cite{schu2} for the details) provides a
connection to the integrable $SU_q(2)$-symmetric $XXZ$ quantum
spin chain with $q=\frac {g_{01}}{g_{10}}$; $H_{XXZ}$ is the
Hamiltonian of the $U_q(su(2))$ invariant quantum spin chain
$H^{QGr}_{XXZ}$ \cite{sal} with anisotropy $\Delta_q$ and with
added non diagonal boundary terms~$B_1$ and~$B_L$
\begin{gather*}
H_{XXZ}=H^{QGr}_{XXZ} + B_1 +B_L \nonumber \\
\phantom{H_{XXZ}}{} =-1/2\sum_{i=1}^{L-1} (\sigma^x_i \sigma^x_{i+1} + \sigma^y_i
\sigma^y_{i+1} - \Delta_q \sigma^z_i \sigma^z_{i+1}
 + h(\sigma^z_{i+1}-\sigma^z_i)+\Delta_q ) + B_1 +B_L.
\end{gather*}
 The transition rates of the ASEP are related to the
parameters $\Delta_q$ and $h$, and the boundary terms in the
following way ($\mu$ is free parameter, irrelevant for the
spectrum)
\begin{gather*}  \Delta_q = -1/2(q^{1/2}+q^{-1/2}), \qquad h=1/2(q^{1/2}-q^{-1/2}),\nonumber  \\
B_1 = \frac {1}{2\sqrt q}\left( \alpha + \gamma + (\alpha -
\gamma) \sigma^z_1 -2\alpha \mu \sigma_1^- -2\gamma \mu^{-1}
\sigma_1^+ \right),\nonumber          \\   B_L=\frac {\left( \beta
+\delta -(\beta-\delta) \sigma^z_L -2\delta \mu
q^{L/2-1/2}\sigma^-_L -2\beta \mu^{-1} q^{-L/2+1/2}\sigma^+_L
\right)}{2\sqrt q}.
\end{gather*}
 For nonequilibrium systems, as
opposed to the ones in equilibrium, the boundary conditions are of
major importance. Emphasizing the dependence of the steady state
behaviour on the boundary rates we study tridiagonal algebras of
the simple exclusion processes which reveal deep algebraic
properties of the latter. In a recent paper \cite{achk} we have
shown that the boundary symmetry of the open partially asymmetric
exclusion process is an Askey--Wilson algebra, the coideal
subalgebra of $U_q(\hat{su}(2))$, which allows for the exact
solvability in the stationary state. From the boundary AW algebra
the tridiagonal  boundary algebra of the ASEP follows through the
natural homomorphism.  The particular values of the structure
constants determine the bulk tridiagonal algebra of the ASEP. It
is generated by the matrices $D_0$ and $D_1$ of the matrix product
ansatz and the def\/ining relations have the form of the level zero
$U_q(\hat{su}(2))$ $q$-Serre relations. This
suggests a framework for analysis of the asymmetric exclusion
process.
 Namely, given the $U_q(\hat{su}(2))$
$R$-matrix operator $R(z_1/z_2)\in {\rm End}_{\textbf{C}} V_{z_1}\otimes
V_{z_2}$, where $V_z$ is the two-dimensional $U_q(\hat{su}(2))$
evaluation module, satisfying the Yang--Baxter equation
\begin{gather*}
R_{12}(z_1/z_2)R_{13}(z_1)R_{23}(z_2)=R_{23}(z_2)R_{13}(z_1)R_{12}(z_1/z_2),
\end{gather*}
 then the transition matrix $\Gamma$ (or
equivalently the Hamiltonian) is written as $H=\sum H_{ii+1}$,
where the two-site  density is obtained as \begin{gather*}
 H_{ii+1}=\frac
{d}{du}PR_{i i+1}\Big| _{u=0} 
\end{gather*} with $P$ the
permutation operator and $z_1/z_2= e^u$. The generators act on the
quantum space by means of the inf\/inite coproduct and the
invariance with respect to the af\/f\/ine $U_q(\hat{su}(2))$
manifests in the property \begin{gather*}
 [H, \Delta^{\infty}(G_k)]=0
\end{gather*} for any of the generators $G_k$ of $U_q(\hat
{su}(2))$. If we introduce for f\/inite chain a boundary of  a~particular form, such as diagonal boundary terms, the symmetry is
reduced to $U_q(su(2))$ and the invariant Hamiltonian \cite{sal}
  $H^{QGr}_{XXZ}$ is known~\cite{schu2} to describe a bulk dif\/fusive
system with ref\/lecting boundaries. In the presence of a general
boundary the symmetry is further reduced to a~boundary symmetry
whose generators are constructed as linear covariant objects with
respect to the bulk quantum group. It turns out that the boundary
symmetry of the asymmetric exclusion process is the AW algebra
whose elements possess  the coproduct properties of two-sided
coideals of the bulk quantum $U_q(su(2))$ symmetry. We have
implemented~\cite{achk} the generators of the AW algebra to
construct an operator valued $K$-matrix, a solution to the
spectral dependent boundary Yang--Baxter equation, also known as a
ref\/lection equation
\begin{gather*}
 R(z_1/z_2)(K(z_1)\otimes
I)R(z_1z_2)(I\otimes K(z_2)) -(I\otimes
K(z_2))R(z_1z_2)(K(z_1)\otimes I)R(z_1/z_2)=0. 
\end{gather*} The
$K(z)$ matrix is the basic ingredient of the inverse scattering
method \cite{skl1} and such a connection points out to exact
solvability beyond the stationary state and description of the
dynamics of the process. This can be achieved by applying the
technique of the quantum inverse scattering method, and in
particular the underlying Bethe Ansatz method to exactly calculate
some  quantities of physical interest, such as the common spectrum
of the commuting conserved quantities.

In the following we f\/irst review the algebra of the asymmetric
exclusion process, both the boundary and the bulk one. The
boundary AW algebra is naturally mapped to a tridiagonal algebra
whose irreducible inf\/inite-dimensional modules are the AW
polynomials as well.  We study the $q=1$ limit of the tridiagonal
algebra and show that it appears to be the boundary symmetry of
the symmetric exclusion process. The structure constants of the
tridiagonal boundary algebra are given in terms of the parameters
of the symmetric exclusion process. From the boundary algebra,
analogously to the case of the ASEP, one obtains, for particular
values of the structure constants,  the bulk tridiagonal algebra
of the symmetric process. We consider the  Askey--Wilson algebra of
the totally asymmetric exclusion process which can be viewed as a
particular $q=0$ limit of the Askey--Wilson algebra. The
consequences of these symmetry properties for the exact
solvability of the simple exclusion model in the stationary state
within the matrix product approach are discussed.

\section{The model within the matrix product approach}

The matrix product approach (MPA) was developed with the aim to
describe the stationary behaviour of many-body systems with
stochastic dynamics. The idea is that the steady state properties
of the ASEP can be obtained exactly in terms of matrices obeying a
quadratic algeb\-ra~\cite{der2, der1}. For a given conf\/iguration
$(s_1, s_2,\dots,s_L)$ the stationary probability is def\/ined by the
expectation value \begin{gather*} P(s)=\frac {\langle w\vert
D_{s_1}D_{s_2}\cdots D_{s_L}
   \vert v\rangle}{Z_L},
\end{gather*}
   where
$D_{s_i}=D_1$, if a site $i=1,2,\dots,L$ is occupied and
$D_{s_i}=D_0$, if a site $i$ is empty. The quantity $Z_L=\langle w
\vert (D_0+D_1)^L \vert v\rangle$ is the normalization factor to
the stationary probability distribution. The operators $D_i$,
$i=0,1$ satisfy the quadratic (bulk) algebra (known as dif\/fusion
algebra~\cite{r})
\begin{gather*}
 D_1D_0-qD_0D_1=x_1D_0-D_1x_0, \qquad
x_0+x_1=0 
\end{gather*} with boundary conditions of the form
\begin{gather*}
  (\beta D_1-\delta D_0)\vert v\rangle  = x_0\vert v\rangle,  \qquad
  \langle w\vert (\alpha D_0 - \gamma D_1) = -\langle w\vert x_1
\end{gather*} and $\langle w\vert v\rangle \neq 0$. The boundary
conditions def\/ine the two vectors $\langle w\vert$ and $\vert
v\rangle$ which enter the expectations values for the stationary
weights. The open model with boundary processes  depends on f\/ive
parameters. These are the bulk probability rate and the four
boundary rates. The partially asymmetric exclusion process (PASEP)
corresponds to $0<q<1$. The limit cases $q=1$ and $q=0$ are the
symmetric exclusion process (SSEP) and the totally asymmetric
exclusion process (TASEP). In the case of TASEP the boundary
processes depend on two boundary rates corresponding to incoming
particles at the left end of the chain and outgoing particles at
the right end.

Given the representations of the quadratic algebra and the
boundary vectors, one can evaluate all the relevant physical
quantities, such as the mean density  at a site $i$, $i=1,\dots,L$,
$\langle s_i\rangle =\frac {\langle
   w\vert (D_0+D_1)^{i-1}D_1(D_0+D_1)^{L-i}
   \vert v\rangle}{Z_L}$,
correlation functions, the current $J$ through a bond between site
$i$ and site $i+1$ which has a very simple form $J=\frac
{Z_{L-1}}{Z_L}$. In most studied examples one uses
inf\/inite-dimensional representations of the quadratic algebra.
Finite-dimensional representations~\cite{ess, ma} have been
considered too and they simplify calculations. Due to a constraint
on the model parameters they def\/ine an invariant subspace of the
inf\/inite matrices and give exact results only on some special
curves of the phase diagram.

Exact results for the ASEP with open boundaries were obtained
within the MPA through the relation of the stationary state to
$q$-Hermite \cite{ev} and Al-Salam--Chihara polynomials \cite{sa}
in the case $\gamma = \delta =0$ and to the Askey--Wilson
polynomials \cite{wa} in the general case. The MPA was readily
generalized to many-species models \cite{ar} and to dynamical MPA~\cite{sti}.

Emphasizing the dependence of the open ASEP in the stationary
state on the boundary rates and the equivalence with the spin
$1/2$ $XXZ$ chain we can represent \cite{an} the boundary
operators in the form \begin{gather*}
 \beta D_1 -\delta D_0  =  A  +x_0\frac {\beta -\delta}{1-q},  \qquad
 \alpha D_0 -\gamma D_1  =  A^* + x_0\frac {\alpha -\gamma}{1-q},
\end{gather*} where $A$, $A^*$ are linear combinations of the
$U_q(\hat{su}(2))$  generators in the evaluation representation
\begin{gather*} A  =  -\frac {x_1\beta}{\sqrt {1-q}}q^{N/2}A_+
 -\frac {x_0\delta}{\sqrt {1-q}}A_-q^{N/2}
 -\frac{x_1\beta q^{1/2}+x_0\delta}{1-q}q^N,\nonumber  \\
A^*  =  +\frac {x_0\alpha}{\sqrt {1-q}}q^{-N/2}A_+ + \frac
{x_1\gamma }{\sqrt {1-q}}A_-q^{-N/2}
 +\frac {x_0 \alpha q^{-1/2} + x_1\gamma }{1-q}q^{-N}.
 \end{gather*} The operators $N$, $A_{\pm}$ generate the $U_q(su(2))$
 algebra ($0<q<1$)
 \begin{gather}
  [N, A_{\pm}]= \pm A_{\pm}, \qquad [A_-, A_+]=\frac
{q^N-q^{-N}}{q^{1/2}-q^{-1/2}} \label{eq12}
\end{gather} with a central
element $Q=A_+A_- -\frac {q^{N-1/2} -q^{-N+
1/2}}{(q^{1/2}-q^{-1/2})^2}$. The pair of operators $A$, $A^*$
satisfy the relations of the boundary  Askey--Wilson algebra
\begin{gather} [[A,A^*]_q,A]_q  =  -\rho A^* -\omega A - \eta,  \qquad
 [A^*,[A,A^*]_q]_q  =  -\rho^*A -\omega A^* -\eta^* \label{eq13}
\end{gather}
with  structure constants
\begin{gather}
  \rho=x_0^2 \beta \delta q^{-1}(q^{1/2}+q^{-1/2})^2, \qquad
  \rho^*=x_0^2 \alpha \gamma q^{-1}(q^{1/2}+q^{-1/2})^2,
\label{eq14}\\
 -\omega=x_0^2(\beta -\delta)(\gamma- \alpha)
    -x_0^2( \beta \gamma +\alpha \delta) (q^{1/2}-q^{-1/2})Q,\nonumber\\
 \eta = q^{1/2}(q^{1/2}+q^{-1/2})x_0^3\left
(\beta \delta ( \gamma
 -\alpha)Q +\frac {(\beta -\delta)(\beta \gamma
 +\alpha \delta)}{q^{1/2} - q^{-1/2}} \right ),\nonumber\\
\eta^* = q^{1/2}(q^{1/2}+q^{-1/2})x_0^3\left (\alpha \gamma (\beta
 -\delta)Q + \frac {( \alpha - \gamma)(\alpha \delta
 +  \beta \gamma)}{q^{1/2}-q^{-1/2}} \right ). \nonumber 
\end{gather} Throughout the text
$[X,Y]_q=q^{1/2}XY-q^{-1/2}YX$. We note the relative dif\/ference of
the factor~$q^{1/2}$ in $\rho$ ($q^{-1/2}$ in $\rho^*$) compared
to \cite{an} which is simply due to a rescaling of the genera\-tors~$A$,~$A^*$. It is important to stress (see next section) that the
Askey--Wilson algebra is def\/ined up to an af\/f\/ine transformation of
rescaling the generators by scalars~$t$,~$t^*$. This property can be
used to even hide the parameters~$\beta \delta$ in~$\rho$ ($\alpha
\gamma$ in $\rho^*$). It is known~\cite{koor, zhe1} that the
inf\/inite-dimensional irreducible modules of the AW algebra are the
AW polynomials depending on four parameters $a$, $b$, $c$, $d$. In the
basic representation one $A^*$ is represented by a diagonal and~$A$ -- by a tridiagonal matrix. The sign of $\rho^*$ (respectively
of $\rho$ in the dual representation) is essential~\cite{zhe1}
for the spectrum of the diagonal operator. Since  for a
representation of the AW algebra we use the $U_q(su(2))$ algebra,
which is the limit case of a $(u,-u)$, $u<0$ algebra,  the spectrum
of the diagonal operator is of the form $\sinh$ (see~\cite{an} for
the details). The boundary conditions of the ASEP uniquely relate
the four parameters of the AW polynomials to the four boundary
rates
 $a=\kappa^*_+(\alpha, \gamma)$, $b=\kappa_+(\beta, \delta)$,
$c=\kappa^*_-(\alpha, \gamma)$, $d=\kappa_-(\beta, \delta)$, where
$\kappa_{\pm}^{(*)}(\nu, \tau)$  is \begin{gather*}
 \kappa_{\pm}^{(*)}(\nu, \tau)= \frac {-(\nu -\tau -(1-q)) \pm
  \sqrt {(\nu -\tau -(1-q))^2 +4\nu \tau}}{2\nu}.
\end{gather*} If we use the $(u,u)$ algebra, the limit case of
which corresponds to $q^N+q^{-N}$ in the nominator of equation~\eqref{eq12},
then the spectrum of the diagonal operator is of the form $\cosh$
and $\kappa_{\pm}^{(*)}(\nu, \tau) \rightarrow
\kappa_{\pm}^{(*)}(\nu, -\tau)$.

The transfer matrix $D_0+D_1$ and each of the boundary operators
generate isomorphic AW algebras~\cite{an}. In the tridiagonal
representation the transfer matrix $D_0+D_1$ satisf\/ies the
three-term recurrence relation of the AW polynomials, which was
explored in~\cite{wa} for the solution of the ASEP in the
stationary state. The exact calculation of all the physical
quantities, such as the current, correlation functions etc, in
terms of the Askey--Wilson polynomials was achieved without any
reference to the AW algebra. The  exact solution in the stationary
state in terms of the  AW polynomials was, in our opinion,
possible due to the appearance of the AW algebra as the boundary
hidden symmetry of the ASEP with general boundary conditions. We
have constructed an AW algebra operator valued $K$-matrix \cite{achk}, a solution to the boundary Yang--Baxter equation. The
relation of the boundary AW algebra to the $K$-matrix, satisfying
the ref\/lection equation, reveals deep algebraic properties of the
ASEP allowing to extend the exact solvability beyond the
stationary state. The boundary Yang--Baxter equation  is the basic
ingredient of the inverse scattering method~\cite{skl1} and one
can use its solutions for description of the dynamics of the
process. With the help of the $K(z)$-matrix one can apply the
technique of the quantum inverse scattering method to construct
the spectral dependent transfer matrix and then implement the
underlying Bethe Ansatz method to exactly calculate some
quantities of physical interest, e.g.\ the exact spectrum of the
transition matrix $\Gamma$, or equivalently the Hamiltonian.

\section{Bulk tridiagonal algebra of the simple exclusion process}

For completeness we f\/irst recall the bulk algebra of the partially asymmetric
exclusion process (PASEP).

{\it The bulk tridiagonal algebra of PASEP}:
The operators $D_0$, $D_1$ of the partially asymmetric exclusion process
and their $q$-commutator $[D_0,D_1]_q$
form a closed linear algebra
\begin{gather*}
[D_0,D_1]_q = D_2,\nonumber \\
[D_1,[D_0,D_1]_q]_q = q^{-1/2}x_1(q^{1/2}-q^{-1/2}) \{D_0,D_1\} \nonumber\\
\phantom{[D_1,[D_0,D_1]_q]_q =}{} -q^{-1}x_1^2D_0
  +q^{-1}x_0x_1D_1 -x_0q^{-1/2}(q^{1/2}-q^{-1/2})D_1^2,\nonumber  \\
[[D_0,D_1]_q,D_0]_q=-x_0q^{-1/2}(q^{1/2}-q^{-1/2})\{D_0,D_1\} \nonumber\\
\phantom{[[D_0,D_1]_q,D_0]_q=}{}- x_0^2q^{-1}D_1
   +x_0x_1q^{-1}D_0 -x_1q^{-1/2}(q^{1/2}-q^{-1/2})D_0^2,
\end{gather*}
 where $\{D_0,D_1\}=D_0D_1+D_1D_0$. The algebra can
equivalently be described as a two-relation algebra for the pair
$D_0$, $D_1$
\begin{gather*}
D_0D_1^2-(q+q^{-1})D_1D_0D_1+D_1^2D_0
 + x_1q^{-1/2}(q^{1/2}-q^{-1/2}) \{D_0,D_1\}\nonumber  \\
\qquad{} =  x_1^2q^{-1}D_0 -x_0x_1q^{-1}D_1
     +x_0q^{-1/2}(q^{1/2}-q^{-1/2})D_1^2,\nonumber   \\
D_0^2D_1-(q+q^{-1})D_0D_1D_0+D_1D_0^2
 - x_0q^{-1/2}(q^{1/2}-q^{-1/2}) \{D_0,D_1\} \nonumber\\
 \qquad {} = x_0^2q^{-1}D_1 -x_0x_1q^{-1}D_0 +
x_1q^{-1/2}(q^{1/2}-q^{-1/2})D_0^2. 
\end{gather*} The above
relations are the well known  Askey--Wilson relations
\begin{gather}
 A^2A^*-(q+q^{-1})AA^*A +A^*A^2  - \gamma (AA^*+A^*A)=
 \rho A^* + \gamma^*A^2 + \omega A  +\eta,\nonumber \\
 A^{*2}A -(q+q^{-1})A^*AA^*+AA^{*2}- \gamma^* (AA^*+A^*A) =
 \rho^* A +\gamma A^{*2} + \omega A^* +\eta^*.
\label{eq20}
\end{gather} The algebra \eqref{eq20} was f\/irst considered in the works
of Zhedanov \cite{zhe1, zhe2} who showed that the Askey--Wilson
polynomials give pairs of inf\/inite-dimensional matrices satisfying
the Askey--Wilson (AW) relations. It is recently discussed in a
more general framework of a tridiagonal algebra \cite{ter1,ter2}, that is an associative algebra with a unit generated  by a (tridiagonal) pair of operators $A$, $A^*$ and def\/ining relations,
obtained by taking the commutator with $A$, $A^*$ in the f\/irst
(second) line of \eqref{eq20} respectively
 \begin{gather*}
 [A, A^2A^*- \beta AA^*A +A^*A^2 - \gamma (AA^*+A^*A)-\rho A^*]= 0,\nonumber  \\
 [A^*, A^{*2}A - \beta A^*AA^*+AA^{*2}- \gamma^* (AA^*+A^*A)-\rho^* A]= 0.
\end{gather*}  In the general case a tridiagonal pair is
determined by the sequence of scalars $\beta$, $\gamma$, $\gamma^*$,
$\rho$, $\rho^*$. (We keep the conventional notations, used in the
literature, for the scalars of a tridiagonal pair; $\beta$ and
$\gamma$ should not be confused with the ASEP boundary rates.)
Tridiagonal pairs have been classif\/ied according to the dependence
on the scalars~\cite{ter1}. The  example which is important for
the present study are the Dolan--Grady relations~\cite{dol} with
$\beta=2$, $\gamma = \gamma^*
=0$, $\rho = k^2$, $\rho^* =k^{*2}$
\begin{gather*} [A,[A,[A,A^*]]]=k^2[A,A^*],\qquad
 [A^*,[A^*,[A^*,A]]]=k^{*2}[A^*,A]. 
 \end{gather*} Tridiagonal
pairs are determined up to an af\/f\/ine transformation
\begin{gather*}
 A \rightarrow tA+c, \qquad A^* \rightarrow t^*A^*+c^*,
\end{gather*}
where $t$, $t^*$, $c$, $c^*$ are some scalars. The af\/f\/ine transformation can
be used to bring a tridiagonal pair in a reduced form with
$\gamma = \gamma ^* =0$.

As seen from the Askey--Wilson relations \eqref{eq20} for the matrices
$D_0$ and $D_1$ we have explicitly
\begin{gather*} \rho=x_1^2q^{-1}, \qquad
\rho^*=x_0^2q^{-1}, \qquad
  \omega=-x_0x_1q^{-1},
\\
\gamma=-x_1q^{-1/2}(q^{1/2}-q^{-1/2}), \qquad
 \gamma^*=x_0q^{-1/2}(q^{1/2}-q^{-1/2})
\end{gather*}
and $\eta = \eta^* =0$.
Besides $\gamma=\gamma^*$, $\rho=\rho^*$ due to $x_0+x_1=0$.
We can now rescale the opera\-tors~$D_0$,~$D_1$
to set $\gamma=\gamma^*=0$. This is achieved with the help of the
transformations
\begin{gather*}
D_0 \rightarrow D_0+\frac {x_0q^{-1/2}}{q^{1/2}-q^{-1/2}},
\qquad
D_1 \rightarrow D_1 -\frac {x_1q^{-1/2}}{q^{1/2}-q^{-1/2}}.
\end{gather*}
However the shift of the generators  amounts to
a tridiagonal pair with sequence of scalars $\beta = -(q+q^{-1})$,
$\gamma=\gamma^*=0$, $\rho=\rho^*=0$.
Thus the operators of the ASEP matrix product ansatz obey the
relations of a tridiagonal algebra \begin{gather} [D_1, D_0D_1^2
-(q+q^{-1})D_1D_0D_1 +D_1^2D_0]=0,\nonumber \\    [D_0, D_1D_0^2
-(q+q^{-1})D_0D_1D_0 +D_0^2D_1]=0, \label{eq27}
\end{gather} which is a
special case of the tridiagonal relations of the ASEP boundary
operators with $\beta =q+q^{-1}$ and $\gamma = \gamma^* = \rho =
\rho^* =0$. These relations are the $q$-Serre relations for the
level zero $U_q(\hat{su}(2))$ adjoint representation. As already
mentioned  we have def\/ined the Askey--Wilson algebra with two
generators through a homomorphism to the quantized af\/f\/ine
$U_q(\hat{su}(2))$~\cite{achk}.  It is important to understand the
role of the parameters $x_0$, $x_1$ in the quadratic algebra of the
open ASEP from the point of view of this homomorphism. These
parameters can be interpreted as the weights associated to the
Cartan generators $H_i$, $i=0,1$ of the $U_q(\hat{su}(2))$
f\/inite-dimensional module with weight space $W_{\nu}=(\nu \in
W\vert q^{H_i}\nu=q^{x_i}\nu)$. The $U_q(\hat{su}(2))$ module $W$
has level $k$ if the central element $q^{H_0+H_1}$ acts on it as
the scalar $q^k$~\cite{jimmi}. Hence, it follows from the bulk
tridiagonal algebra \eqref{eq27} that the matrices $D_0$, $D_1$ of the
quadratic bulk algebra of the open ASEP with $x_0+x_1$ correspond
to the two-dimensional level $0$ $U_q(\hat{su}(2))$ adjoint module.

The bulk Askey--Wilson algebra of the symmetric exclusion process
follows immediately as the limit $q\rightarrow 1$ of the bulk
tridiagonal algebra of the  asymmetric exclusion process. Hence we
have:

\begin{proposition}\label{proposition1}
 The operators $D_0$, $D_1$ of the symmetric
simple exclusion process and their commutator $[D_0,D_1]$ form a
closed linear algebra
\begin{gather*} [D_0,D_1] = D_2,\qquad
[D_1,[D_0,D_1]] = x_1 \{D_0,D_1\} -x_1^2D_0
  +x_0x_1D_1 -x_0D_1^2,\nonumber  \\
[[D_0,D_1],D_0] = -x_0\{D_0,D_1\}
  -x_0^2D_1
   +x_0x_1D_0 -x_1D_0^2.
\end{gather*}
\end{proposition}

The proposition can be independently verif\/ied by
directly using the explicit form of the MPA quadratic relation for
the case of symmetric dif\/fusion. The algebra can  equivalently be
described as a two-relation algebra for the pair $D_0$, $D_1$
\begin{gather*}
D_0D_1^2-2D_1D_0D_1+D_1^2D_0 +x_1\{D_0,D_1\}
    = x_1^2D_0 -x_0x_1D_1
     +x_0D_1^2,\nonumber   \\
D_0^2D_1-2D_0D_1D_0+D_1D_0^2 -x_0\{D_0,D_1\}
 =   x_0^2D_1 -x_0x_1D_0 +
x_1D_0^2.
\end{gather*}
As seen from the explicit form of the algebra, the matrices $D_0$, $D_1$ of the symmetric
exclusion process satisfy the Askey--Wilson relations with the sequence of
scalars
\begin{gather*}
\beta=2, \qquad \rho=x_1^2, \qquad \rho^*=x_0^2, \qquad \omega=-x_0x_1,
  \qquad \gamma=-x_1, \qquad \gamma^*= -x_0.
\end{gather*} We can transform the generators $D_0$, $D_1$ by the
af\/f\/ine shifts
\begin{gather*}
 D_0 \rightarrow D_0+x_0, \qquad D_1
\rightarrow D_1 -x_1. 
\end{gather*} By a straightforward
calculation we again observe  the same property as in the case of
the partially asymmetric process. Namely, the shifts of the
generators amount to a tridiagonal pair with the sequence of
scalars $\gamma=\gamma^*=0$, $\rho=\rho^*=0$.
Thus the operators of the symmetric matrix product ansatz obey the
relations of a tridiagonal algebra
\begin{gather*}
 [D_1, D_0D_1^2 -2D_1D_0D_1
+D_1^2D_0]=0, \qquad [D_0, D_1D_0^2 -2D_0D_1D_0 +D_0^2D_1]=0,
\end{gather*}
 which are the Dolan--Grady relations with
$k=k^*=0$. In the next section we show that the bulk tridiagonal
relations  are the  special case of the tridiagonal relations
obeyed by the boundary operators of the symmetric process.

\section{Boundary algebra of the symmetric exclusion process}

We consider the symmetric simple exclusion process (SSEP) with
most general boundary conditions of incoming and outgoing particle
at both ends of the chain. Within the matrix product ansatz the
quadratic bulk algebra of the SSEP is the $q=1$ limit of the
deformed quadratic algebra of the ASEP. In the previous section we
derived the bulk tridiagonal algebra of the symmetric process and
it turned to be the $q=1$ limit of the bulk tridiagonal algebra of
the asymmetric process.

We are going to obtain the boundary algebra of the symmetric
process as the $q=1$ limit of the tridiagonal boundary algebra of
the asymmetric simple exclusion process. We start with the
observation about the  natural homomorphism of the tridiagonal
algebra (TA) generated by the pair $A$, $A^*$ and the Askey--Wilson
algebra (AW) of the ASEP def\/ined by equation \eqref{eq13}, namely $TA
\rightarrow AW$. As already mentioned, this is readily verif\/ied by
taking the commutator with $A$, respectively $A^*$ of the f\/irst
line, respectively the second line
of equation~\eqref{eq13}, which gives
\begin{gather}
 [A,[A,[A,A^*]_q,]_{q^{-1}}]]= \rho [A,A^*],\qquad
 [A^*,[A^*[A^*,A]_q]_{q^{-1}}] = \rho^*[A^*,A]  \label{eq33}
\end{gather}
with $\rho$, $\rho^*$ depending on the f\/ive parameters of the ASEP,
as given by \eqref{eq14}. The opera\-tors~$A$,~$A^*$ were introduced as the
shifted boundary operators of the ASEP. The limit $q=1$ of the
tridiagonal algebra \eqref{eq33} provides a way to determine the boundary
algebra of the SSEP.

 The  boundary operators of the SSEP
can be represented in the form (i.e.\ linear combinations  in terms
of the level zero af\/f\/ine $su(2)$ generators) \begin{gather} \beta D_1
-\delta D_0 = -x_1\beta A_+
 -x_0\delta A_-
  -(x_1\beta +x_0\delta)N
 - x_1\beta -x_0\delta,\nonumber     \\
\alpha D_0 - \gamma D_1 =
 x_0\alpha A_+ + x_1\gamma A_-
+ (x_0 \alpha  + x_1\gamma )N  + x_0\alpha + x_1\gamma, \label{eq34}
\end{gather} where $A_{\pm}$, $N$ are corresponding operators in the limit
$q\rightarrow 1$ of equation~\eqref{eq12}.  We separate the shift parts from the
boundary operators. Denoting the corresponding rest operator parts
by~$A$ and~$A^*$ we write the left and right boundary operators in
the form \begin{gather} \beta D_1 -\delta D_0  =  A - x_1\beta-x_0\delta,\qquad
 \alpha D_0 - \gamma D_1  =  A^* + x_0\alpha + x_1\gamma.
\label{eq35}
\end{gather}
Then we have the following

\begin{proposition}\label{proposition2}
The operators $A$ and $A^*$ defined by
the corresponding shifts of the boundary operators of the
open symmetric exclusion process
\begin{gather*}
 A =
 \beta D_1 -\delta D_0 + (x_1\beta+x_0\delta),\qquad
 A^* =
 \alpha D_0 - \gamma D_1- (x_0\alpha + x_1\gamma)
\end{gather*}
and their commutator
\begin{gather*}
 [A,A^*]=AA^*-A^*A
\end{gather*}
 form a closed linear algebra, the boundary
tridiagonal algebra of the SSEP
 \begin{gather}
  [A,[A,[A,A^*]]]= \rho [A,A^*],\qquad
 [A^*,[A^*,[A^*,A]]] = \rho^*[A^*,A], \label{eq38}
 \end{gather}
  where the
structure constants are given by
\begin{gather}
 -\rho=x_0x_1 \beta \delta , \qquad
 -\rho^*=x_0x_1 \alpha \gamma
\label{eq39}
\end{gather}
\end{proposition}

  This proposition is straightforward to verify by
taking the $q\rightarrow 1$ limit in the chain of homomorphisms
$TD \rightarrow AW \rightarrow U_q(\hat{su}(2))$ or independently
by directly using equation~\eqref{eq34}.

As readily seen from the def\/inition~\eqref{eq35} the (shifted) boundary
operators of the symmetric exclusion process obeying the  algebra
\eqref{eq38} form a tridiagonal pair with $\beta = 2$, $\gamma = \gamma^*
=0$, and~$\rho$,~$\rho^*$ given by~\eqref{eq39}. The tridiagonal boundary
algebra of the symmetric process is the limit $q=1$ of the
deformed  boundary algebra of the ASEP as the irreducible modules
of the algebra in the symmetric model, i.e.\ the Wilson
polynomials, are the $q=1$ limit of the Askey--Wilson polynomials.
The important properties of the deformed Askey--Wilson algebra
remain valid in the proper limit $q\rightarrow 1$. Stated more
precisely which will correspond to the historical development of
these algebras, for generic $q$ the deformed Askey--Wilson algebra
is the $q$-generalization of the Onsager algebra in the form of
the Dolan--Grady relations. Its irreducible modules, i.e.\ the
Askey--Wilson polynomials are $q$-counterpart of the Wilson
polynomials. For applications, howe\-ver one can even use the limit
cases of the Wilson polynomials. These are the continuous Hahn and
dual Hahn polynomials (see~\cite{koe} and~\cite{ga} for details
on these polynomials). In both cases there exist limiting
procedures to further obtain the Meixner--Pollaczek polynomials
from which the Laguerre polynomials can be obtained. Let $P_n^{\mu
}(x, \phi )$ denote the $n$-th Meixner--Pollaczek
\begin{gather*}
 P_n^{\mu}(x,
\phi)=\frac {(2\mu)_n}{n!}e^{in\phi}\,
 {}_2\!F_1 \left ( \begin {array}{c}
   -n, \mu+ix \\
     2\mu  \end{array} \vert 1-e^{-2i\phi} \right ).
\end{gather*}
The Laguerre polynomials can be obtained by the substitution
$\mu=1/2(\lambda+1)$, $x\rightarrow -1/2\phi^{-1}(x)$ and letting
$\phi \rightarrow 0$
\begin{gather*}
\lim_{\phi \rightarrow 0}
P_n^{1/2\mu +1/2}\left (-\frac {x}{2\phi};\phi \right )= L_n^{(\lambda)}(x)
\end{gather*}
By def\/inition the Laguerre polynomials have the form
\begin{gather*}
L_n^{(\lambda)}(x)= \frac {(\lambda +1)_n}{n!}\,
 {}_1\!F_1 \left ( \begin {array}{c}
   -n  \\
     \lambda +1  \end{array} \vert x \right )
\end{gather*} with orthogonality condition
\begin{gather*}
 \int_0^{\infty}
e^{-x}x^{\lambda}L_m^{\lambda}L_n^{\lambda}dx= \frac {\Gamma
(n+\lambda+1)}{n!}\delta _{mn}, \qquad \lambda > -1 
\end{gather*} and recurrence relation
\begin{gather*}
 (n+1)L_{n+1}^{(\lambda)}(x)-
(2n+\lambda +1-x)L_n^{(\lambda)}(x) +(n+\lambda
)L_{n-1}^{(\lambda)}(x)=0, 
\end{gather*} where
$L_{-1}^{(\lambda)}(x)=0$ and $L_0^{(\lambda)}(x)=1$. One can
identify \begin{gather*}
 \lambda = \frac {\alpha +\beta +\gamma +\delta
}{(\alpha +\gamma )(\beta +\delta )}-1. 
\end{gather*}
 Denoting
\begin{gather*}
 l_n(x)=(-1)^n \left ( \frac {n!\Gamma (\lambda +1)}{\Gamma (\lambda +n+1)} \right )^{1/2}
 L_n^{(\lambda)}(x)
\end{gather*}
we rewrite the orthogonality condition in the form
\begin{gather*}
1= \frac {1}{\Gamma (\lambda +1)}
   \int_0^{\infty} e^{-x}x^{\lambda}\vert l(x)\rangle \langle l(x) \vert dx.
\end{gather*}
 The vectors  $\vert l(x)\rangle=
(l_0(x),l_1(0),\dots)^t$  and $\langle l(x)\vert=
(l_0(x),l_1(0),\dots)$ form the basis for the tridiagonal and the
diagonal representation of the generators (and for the dual one).
As it was proved to be the case for the ASEP~\cite{an}, each
boundary operator of the symmetric process, together with the
transfer matrix operator $D_0+D_1$, forms an isomorphic AW algebra
whose tridiagonal representation follows from the three-term
recurrence relation, consistent with the orthogonality condition
\begin{gather}
 (D_0+D_1)\vert l(x)\rangle =x\vert l(x)\rangle, \qquad
\langle l(x)\vert (D_0+D_1)=\langle l(x)\vert x. \label{eq48}
\end{gather}
These properties can be used to exactly calculate the physical
quantities~\cite{wa} in the stationary state. The results \cite{schu1} for the partition function $Z_l$ and the current   are
reconstructed \begin{gather*}
  Z_L= \frac {\Gamma (\lambda +L+1)}{\Gamma (\lambda +1)},\qquad
    J=\frac {1}{\lambda +L}.
\end{gather*}
 The one point function \cite{schu1}
 \begin{gather*} \langle
\tau_i \rangle = \frac {\alpha}{\alpha+\gamma} -\frac{1}{\lambda
+L} \frac {\alpha \beta - \gamma \delta}{(\alpha +\gamma )(\beta
+\delta )} \left (\frac {1}{\alpha+\gamma} +i-1 \right ), \qquad
i=1,2,\dots,L 
\end{gather*} shows that the particle density has a
linear prof\/ile. Without reference to the AW algebra the eigenvalue
equation~\eqref{eq48} was used in \cite{wa} to calculate the physical
quantities of the symmetric exclusion process, as known from the
matrix product approach. In our opinion it is the boundary
tridiagonal algebra of the symmetric exclusion process that allows
for the exact solvability of the symmetric process in the
stationary state and leads to a generalization of the matrix
product method.

\section{Nonlocal conserved charges\\ of the symmetric exclusion process}

In the previous section we have shown that the boundary symmetry
of the symmetric exclusion process is the tridiagonal  algebra
\eqref{eq38} with the sequence of scalars
 $\beta=2$, $\rho$, $\rho^*$.
 This algebra is the $q=1$ limit of the tridiagonal
algebra mapped through the natural homomorphism to the Zhedanov
algebra $AW(3)$ def\/ined for $0<q<1$. The exact calculation of the
physical quantities in the case of the symmetric exclusion process
has been obtained in terms of the Laguerre polynomials
implementing~\cite{wa} the three-term recurrence relation to
def\/ine  the tridiagonal representation of the transfer matrix
$D_0+D_1$.  As we pointed out, the ultimate relation of the exact
solution in the stationary state  to the Laguerre polynomials was
possible due to the $q=1$ boundary hidden symmetry of the SSEP
with general boundary conditions.

The def\/ining relations \eqref{eq38} of the $q=1$ tridiagonal  boundary
algebra can (with $\rho$, $\rho^*$ expressed in terms of the
boundary rates according to~\eqref{eq39}) are the well-known Dolan--Grady
relations for the shifted boundary operators. The importance of
the Dolan--Grady relations is that given a~self-dual Hamiltonian
\begin{gather*}
 H=fA+f^*A^*,
\end{gather*}
where   $f$, $f^*$  are some coupling constants, and
$A$, $A^*$ satisfy the relations \eqref{eq38}, then one can construct (an
inf\/inite set of) conserved commuting charges
\begin{gather*}
 Q_{2n}=f(R_{2n}-\tilde {R} _{2n-2})+f^*(\tilde {R} _{2n} - R_{2n-2})
\end{gather*}
 in terms of the quantities
 \begin{gather}
  R_{2n}= -\frac
{2}{\rho}[A[A^*,R_{2n-2}]]-\tilde {R} _{2n-2}\label{eq54}
\end{gather}
with
 $R_0\equiv A$, 
$Q_0\equiv H$. 
  In the
case of the symmetric exclusion process from the boundary
conditions we have
\begin{gather*}
 \langle w \vert A-(x_1\beta+x_0\delta)
\vert v \rangle  =  x_0 \langle w \vert v \rangle , \qquad \langle
w \vert  A^* + (x_0\alpha + x_1\gamma) \vert v \rangle  =  -x_1
\langle w \vert v . 
\end{gather*}
 Hence \begin{gather*}
  \langle w \vert x_0(
A^* + (x_0\alpha + x_1\gamma)) + x_1(A - (x_0\beta+x_1\delta))
\vert v \rangle =0 
\end{gather*} and we can interpret the quantity
\begin{gather*}
 x_0( A^* + (x_0\alpha + x_1\gamma)) + x_1(A - (x_0\beta+x_1\delta))
 \end{gather*}
  as the Hamiltonian
of the symmetric simple exclusion process in the auxiliary space.
As we know the generators of the tridiagonal algebra are
determined up to shift transformations. In view of this property
it seems more convenient to consider the shifted Hamiltonian
\begin{gather*}
 H_s=x_0A^* + x_1A,
\end{gather*} which is self-dual, if we def\/ine $x_1^*=x_0$. Then we can
straightforward apply the prescription of Dolan and Grady to
obtain the  conserved nonlocal charges. Taking into account the
shifts we f\/ind the result
\begin{gather*} R_0=A-x_1\beta -x_0\delta, \qquad 
 \tilde {R}_0=A^*+x_0\alpha +x_1\gamma, \\ 
R_2=-\frac
{2}{\rho}[A-x_1\beta-x_0\delta,[A^*+x_0\alpha+x_1\gamma,
A-x_1\beta -x_0\delta]]-A^*+x_0\alpha+x_1\gamma 
\end{gather*}
 and
so on, according to formula \eqref{eq54}. The expressions
$A-x_1\beta-x_0\delta$ and $A^*+x_0\alpha+x_1\gamma$ are the right
and left boundary operators \begin{gather*}
 B^R=\beta D_1-\delta D_0, \qquad
B^L= \alpha D_1-\gamma D_0 
\end{gather*}
 respectively which
acquire a very important physical meaning. The boundary operators
satisfying \begin{gather}
 [B^R,[B^R,[B^R,B^L]]]= -x_0x_1\beta \delta [B^R,B^L],  \nonumber\\
  [B^L,[B^L,[B^L,B^R]]] = -x_0x_1\alpha \gamma [B^L,B^R]
\label{eq64}
\end{gather}
 are the nonlocal conserved charges of the symmetric
exclusion process with the help of which the (inf\/inite) set of
conserved quantities of the process are constructed. By properly
rescaling the operators one achieves  equal coef\/f\/icient factors on
the RHS of equations~\eqref{eq64}. We note that quantum integrals of motion for
the $XXX$ Heisenberg inf\/inite chain (known to be related to the
symmetric exclusion process)
 were f\/irst obtained in~\cite{gra}. For the open symmetric exclusion
 process the existence of  nonlocal conserved quantities will result in
the exact solvability of the system beyond the stationary state
due to the boundary hidden tridiagonal symmetry of the symmetric
exclusion process.

\section[Askey-Wilson algebra of the totally asymmetric exclusion process]{Askey--Wilson algebra of the totally asymmetric\\ exclusion process}

The tridiagonal and the AW algebra of the totally asymmetric
exclusion process cannot be obtained directly as the limit $q=0$
of the partially asymmetric process. The procedure is more
involved. We derive this algebra from the quadratic algebra of the
totally asymmetric process.

We start with the quadratic algebra
\begin{gather}
D_1D_0=D_1+D_0.
\label{eq65}
\end{gather}
From this algebra the following relations follow
\begin{gather}
D_1D_0D_1=D_1^2+D_0D_1, \qquad  D_0D_1D_0=D_0D_1 +D_0^2
\label{eq66}
\end{gather}
and
\begin{gather}
D_1^2D_0=D_1^2+D_1+D_0, \qquad D_1D_0^2=D_1+D_0+D_0^2,
\label{eq67}
\end{gather}
which can alternatively be written as
\begin{gather}
D_1D_0D_1-D_1^2D_0  =  [D_0,D_1],\qquad
D_0D_1D_0-D_1D_0^2  =  [D_0,D_1]. \label{eq68}
\end{gather}
 The 
 LHS of \eqref{eq68} are respectively \begin{gather}
  [D_1D_0, D_1], \qquad  [D_0, D_1D_0].
\label{eq69}
\end{gather}
Hence we have
\begin{gather}
 D_1[D_0,D_1]=[D_0,D_1] , \qquad  [D_0,D_1]D_0=[D_0,D_1].
\label{eq70}
\end{gather}
Examples of  matrices obeying the above relations
are given by the equation~\eqref{eq33} and (36), (38) in~\cite{der2}. We can
shift the operators $D_0$, $D_1$ by~$1$ (or respectively by the
constants $c_0=a$, $c_1=b$) \begin{gather*}
 D_0\rightarrow D_0+1,
\qquad D_1\rightarrow D_1+1. 
\end{gather*}
 Then we can
write subsequently 
 $D_1D_0=Z$, 
 where either $Z=1$
or $Z=ab$. (In the case of the totally asymmetric exclusion
process within the matrix product approach $ab=\alpha \beta$,
where~$\alpha$ and~$\beta$ are the probability rates for the
particles to be added and removed at both sides of the linear
chain.) Hence \begin{gather}
 D_1D_0D_1=ZD_1, \qquad  D_0D_1D_0=ZD_0,
\label{eq73} \\ D_1^2D_0=ZD_1, \qquad  D_1D_0^2=ZD_0.
\label{eq74}
\end{gather}
 Relations \eqref{eq68} for the shifted generators  become
\begin{gather}
D_1D_0D_1-D_1^2D_0  =  0,\qquad
D_0D_1D_0-D_1D_0^2  =  0 \label{eq75}
\end{gather}
and consequently
\begin{gather}
D_1[D_0,D_1]= 0, \qquad  [D_0,D_1]D_0=0. \label{eq76}
\end{gather} We
need to emphasize that  multiplying in equations~\eqref{eq75}, the f\/irst equation by
$D_1$ from the right and the second equation by $D_0$ from the left,
we obtain \begin{gather}
D_1D_0D_1^2-D_1^2D_0D_1  =  0,\qquad
D_0^2D_1D_0-D_0D_1D_0^2  =  0, \label{eq77}
\end{gather} which  def\/ine the
$q=0$ limit of the $q$-Serre relations of $U_q (\hat{sl}(2))$,
i.e.\ the $q=0$ limit of the level zero $U_q (\hat{sl}(2))$ adjoint
representation. This is consistent with the def\/inition of the
Askey--Wilson algebra such, that it  yields a deformation of the
level zero $U_q (\hat{sl}(2))$ $q$-Serre relations (see \cite{achk} for details). However, we have now the additional relation
\begin{gather*}
 ([D_0,D_1])^2=Z[D_0,D_1].
\end{gather*}
 It is important to emphasize that the matrices
obeying \eqref{eq65}--\eqref{eq70} and \eqref{eq73}--\eqref{eq76} are upper bi\-diagonal and lower
bi\-diagonal. It is known that the basic representation of the AW
algebra can be equivalently considered as a representation on the
space of $c$-number sequences where one of the generators acts as
a tridiagonal operator, while the other generator acts as a
diagonal operator.  In our case the analogues of the diagonal and
tridiagonal matrices in the $q=0$ limit of the basic
representation are given by $[D_0,D_1]$ and $D_1+D_0$
respectively.

\begin{proposition}\label{proposition3}
The $q=0$ AW algebra, as following from the quadratic algebra \eqref{eq65}, depending on only two
constants $a$, $b$ (i.e.\ $e_1=a+b$, $e_2=ab$) in the basic
representation, is def\/ined by
\begin{gather*}
 D_1D_0=e_2, \qquad
D_1[D_0,D_1]= 0, \qquad  [D_0,D_1]D_0=0, \\ 
[D_0,D_1](D_0+D_1)[D_0,D_1]=0, \qquad
([D_0,D_1])^2=e_2[D_0,D_1]. 
\end{gather*} The matrix
$D_0+D_1$, denoted hereafter $D$,  plays the role of the
tridiagonal operator, and $[D_0,D_1]$, denoted hereafter $D^*$,~--
the role of the diagonal operator, in the basic
representation.

Thus we have \begin{gather*}
 [D_0,D_1](D_0+D_1)^2[D_0,D_1]=e_2^2[D_0,D_1],\\ 
 [D_0,D_1](D_0+D_1)[D_0,D_1]=0 , \qquad 
([D_0,D_1])^2=e_2[D_0,D_1]. 
\end{gather*}
We can now shift the matrix $(D_0+D_1) \equiv D\rightarrow D+ a+b$ to obtain
a tridiagonal matrix with entries on the main diagonal too.
\end{proposition}

\begin{definition}\label{definition1}
 The $q=0$ limit of the AW algebra depending on
only two constants $a$, $b$, where
 $e_1=a+b$, 
$e_2=ab$, 
 is generated by a tridiagonal operator $D$
and a diagonal operator $D^*$ with def\/ining relations in the basic
representation
\begin{gather}
 D^*DD^*=e_1e_2D^*, \qquad
(D^*)^2=e_2D^*. \label{eq85}
\end{gather}
\end{definition}

One can alternatively consider def\/ining relations for $q=0$ limit
of the AW algebra in a~representation associated with the $q=0$
limit of the level zero adjoint $U_q (\hat{sl}(2))$ (equation~\eqref{eq77}).

\begin{definition}\label{definition2}
The $q=0$ limit of the AW algebra, generated
by  upper diagonal and lower diagonal matrices $D_1$ and $D_0$ and
depending on only two constants $a$, $b$,  is def\/ined by
\begin{gather}
D_1D_0D_1=abD_1, \qquad  D_0D_1D_0=abD_0. \label{eq86}
\end{gather}

The boundary tridiagonal AW algebra of the totally asymmetric
exclusion process corresponds to the values $a=\alpha$, $b=\beta$
while the bulk  AW algebra is obtained for $a=b=1$.
\end{definition}

 We note that the two representation dependent def\/initions of the algebras
considered above have in common  the f\/irst relation in formula
\eqref{eq85} and both relations in \eqref{eq86},  which can be unif\/ied~as
\begin{gather}
ABA=\tilde c (a,b) A, \label{eq88}
\end{gather} where $A$, $B$ are the
generators of these algebras and $\tilde c (a, b)$ is a constant
depending on the parameters $a$, $b$. We can multiply equation~\eqref{eq88} by $B$
subsequently from the left and from the right to obtain
\begin{gather}
ABAB-BABA = 0 \label{eq89}
\end{gather}
 if $\tilde c (a, b)=0$ or
 \begin{gather}
 ABAB-BABA = \tilde c(a,b)(AB-BA) \label{eq90}
 \end{gather}
  if $\tilde
c(a,b)\neq 0$. We can consider equation~\eqref{eq90} as def\/ining an alternative
$q=0$ limit algebra, obtained through the natural homomorphism to
the  $q=0$ limit Askey--Wilson algebra, with structure constant
$\tilde c(a,b)$. These relations can be very useful for
applications. Namely, with the additional condition $A^2=A$, which
is a Hecke type relation equations~\eqref{eq89} and~\eqref{eq90} have the form of a
ref\/lection equation and a modif\/ied boundary Yang--Mills equation,
respectively. The interpretation of equations~\eqref{eq89} and~\eqref{eq90} as
ref\/lection equations should be  associated with the proper
$R$-matrix operator depending on a parameter $t\neq q$.

Thus for the totally asymmetric exclusion process we obtain  the
bulk algebra
\begin{gather*}
 D_1D_0D_1D_0-D_0D_1D_0D_1=D_1D_0-D_0D_1
\end{gather*}
 and the boundary algebra generated by the right $D^R$
and left $D^*L$ boundary operators
\begin{gather*}
 D^R=\beta D_1, \qquad
D^L=\alpha D_0 
\end{gather*}
 subject to the relations
 \begin{gather}
 D^RD^LD^RD^L-D^LD^RD^LD^R=\alpha \beta (D^RD^L-D^LD^R).
\label{eq93}
\end{gather}
It is worth studying the connection of the exact
solvability of the totally asymmetric exclusion process in the
stationary state to the integrability properties based on the
boundary Yang--Baxter equation. The consequences of equation~\eqref{eq93} will be
to naturally extend the exact solvability beyond the stationary
state. In~\cite{deg} Bethe ansatz equations were derived and the
exact spectrum of the transfer matrix of the totally asymmetric
exclusion process was analyzed. In our opinion  the boundary AW
algebra  is the hidden symmetry behind the Bethe ansatz
solvability of the totally asymmetric exclusion process.

\section{Conclusion}
We have studied the symmetry properties of the simple exclusion
process. Our consideration extends the previously obtained results
about the AW algebra (known as the $AW(3,q))$ Zhedanov algebra) to
be the boundary hidden symmetry of the asymmetric exclusion
process. We have shown that the boundary symmetry algebra of the
symmetric simple exclusion process is the $q=1$ limit of the
boundary tridiagonal algebra of the partially asymmetric exclusion
process. It is this algebra, to be denoted $TA(q=1)$, that allows
for the exact solvability of the symmetric model in the stationary
state. The def\/ining relations of the algebra $TA(q=1)$ are the
known Dolan--Grady relations. The consequence of this fact is that
one can def\/ine conserved nonlocal charges of the symmetric
process, which
allows for extending the exact solvability beyond the stationary
state. We have also derived the bulk and boundary AW algebra of
the totally asymmetric exclusion process, which can be viewed as a
particular case of the algebra $AW(3, q=0)$.  The def\/ining
relation of the  boundary algebra of the totally asymmetric
process leads to a boundary Yang--Baxter equation which might be
interpreted again as the deep algebraic reason behind the exact
solvability.

\subsection*{Acknowledgments}
A CEI grant for participation
in the Seventh International Conference ``Symmetry in Nonlinear Mathematical Physics''
is gratefully acknowledged.
The author would like to thank the organi\-zers for the invitation
to participate the conference Symmetry-2007 and for the warm
atmosphere during the stay in Kyiv.

\pdfbookmark[1]{References}{ref}
\LastPageEnding

\end{document}